\documentclass[prl,twocolumn,floatix,amssymb,showpacs,amsmath,superscriptaddress]{revtex4-1}
\usepackage{graphicx}
\usepackage[usenames,dvipsnames]{color}
\usepackage{epsfig,color}
\usepackage{amsmath,amssymb}
\usepackage{mathptmx}
\usepackage[T1]{fontenc}
\usepackage{verbatim}
\usepackage{float}
\usepackage{units}
\usepackage{amsmath}
\usepackage{graphicx}
\usepackage[colorlinks = true,
            linkcolor = NavyBlue,
            urlcolor  = NavyBlue,
            citecolor = NavyBlue,
            anchorcolor = blue]{hyperref}
\usepackage{epstopdf}

\makeatletter

 \@ifundefined{textcolor}{}
 {
   \definecolor{BLACK}{gray}{0}
   \definecolor{WHITE}{gray}{1}
   \definecolor{RED}{rgb}{1,0,0}
   \definecolor{GREEN}{rgb}{0,1,0}
   \definecolor{BLUE}{rgb}{0,0,1}
   \definecolor{CYAN}{cmyk}{1,0,0,0}
   \definecolor{MAGENTA}{cmyk}{0,1,0,0}
   \definecolor{YELLOW}{cmyk}{0,0,1,0}
 }

\makeatother

\begin{document}

\title{Efficient Quantum Algorithm for Computing \(n\)-time Correlation Functions}
\author{J. S. Pedernales}
\affiliation{Department of Physical Chemistry, University of the Basque Country UPV/EHU, Apartado  644, 48080 Bilbao, Spain}
\author{R. Di Candia}
\affiliation{Department of Physical Chemistry, University of the Basque Country UPV/EHU, Apartado   644, 48080 Bilbao, Spain}
\author{I. L. Egusquiza}
\affiliation{Department of Theoretical Physics and History of Science,~University~of the Basque~Country UPV/EHU,~Apartado \ 644, 48080 Bilbao, Spain}
\author{J. Casanova}
\affiliation{Department of Physical Chemistry, University of the Basque Country UPV/EHU, Apartado   644, 48080 Bilbao, Spain}
\author{E. Solano}
\affiliation{Department of Physical Chemistry, University of the Basque Country UPV/EHU, Apartado   644, 48080 Bilbao, Spain}
\affiliation{IKERBASQUE, Basque Foundation for Science, Alameda Urquijo 36, 48011 Bilbao, Spain}

\begin{abstract}
We propose a method for computing \(n\)-time correlation functions of arbitrary spinorial, fermionic, and bosonic operators, consisting of an efficient quantum algorithm that encodes these correlations in an initially added ancillary qubit for probe and control tasks. For spinorial and fermionic systems, the reconstruction of arbitrary \(n\)-time correlation functions requires the measurement of two ancilla observables, while for bosonic variables time derivatives of the same observables are needed. Finally, we provide examples applicable to different quantum platforms in the frame of the linear response theory.
\end{abstract}


\maketitle

Quantum mechanics is a recipe for computing probability distributions of measurement outcomes in given experiments, typically, at the microscopic scale~\cite{Sakurai}. In development since the early years of the twentieth century, quantum mechanics has allowed us to describe the most fundamental  properties of light and matter, such  as quantum superposition and entanglement~\cite{Horodecki1}, or the behavior of  elementary particles emerging from scattering processes~\cite{Peskin}.  More recently, with the advent  of modern quantum technologies~\cite{LeibfriedEtAl, Devoret13, Bloch05, Obrien09},   quantum mechanics has become the roadmap for the design of  computational protocols and  simulations of physical systems beyond the capabilities of  classical devices~\cite{Feynman82, Lloyd96}. In this renewed view, proof-of-principle experiments have implemented quantum simulations with the promise of an exponential speed-up in the information processing~\cite{Greiner02, Friedenauer08, Kim09, Kim10,Gerritsma1, Gerritsma2, Britton12}. 

According to quantum theory, all information about a system, its stationary states and its evolution, is encoded in the Hamiltonian. Nonetheless, for most cases, the extraction of this information may not be straightforward~\cite{Kitaev95, Abrams99}. Therefore, alternative strategies are needed to identify and obtain  measurable quantities that  characterize the relevant physical information~\cite{DiCandia13}. A case of particular importance is given by  response functions and susceptibilities, which in  the linear response theory are computed in terms of two-time correlation functions ~\cite{Kubo57, Zwanzig65,Forster}. For example, the knowledge of two-time correlation functions of  the form $\langle \Psi | A(t) B(0) | \Psi \rangle$, stemming from perturbation theory, provides us with a microscopic derivation of useful quantities such as conductivity and magnetization~\cite{FetterWalecka}. The reconstruction of time-correlation functions, however, need not be trivial at all, and could profit from quantum algorithm and simulation protocols for their determination. The computation of time correlation functions for propagating signals is at the heart of quantum optical methods~\cite{MandelWolf}, including the case of propagating quantum microwaves~\cite{Bozyigit,DualPathExp,DualPathTheory}. However, these methods are not necessarily easy to export to the case of spinorial, fermionic and bosonic degrees of freedom of massive particles. In this sense, recent methods have been proposed for the case of two-time correlation functions associated to specific dynamics in optical lattices~\cite{Lukin}, as well as in setups where post-selection and cloning methods are available~\cite{Vedral}.  On the other hand, in quantum computer science the SWAP test~\cite{Buhrman01} represents a standard way to access $n$-time correlation functions if a quantum register is available that is, at least,  able to store two copies of a  state,  and to perform a generalized-controlled swap gate~\cite{Wilmott12}. However, this could be demanding if the system of interest is large, for example, for an $N$-qubit system the SWAP test requires the quantum control of a system of more than $2N$ qubits. Another possibility corresponds to the Hadamard test~\cite{Somma02} that requires controlled-time evolutions. The latter is demanding if the dynamics of interest involve many-body or time-dependent Hamiltonians. In contrast to this, here we  present a protocol that exploits the natural evolution of the system and that requires the addition of only one qubit.

Let us thus consider a two-time correlation function $\langle A(t) B(0) \rangle$ where $A(t)= U^{\dagger}(t) A(0) U(t)$, $U(t)$ being a given unitary operator, while $A(0)$ and $B(0)$ are both Hermitian. Remark that, generically,  $A(t)B(0)$ will not be Hermitian. However, one can always construct two self-adjoint operators $C(t) = \frac{1}{2} \{ A(t), B(0) \}$ and $D(t) = (1/2i) [ A(t) , B(0) ]$ such that $\langle A(t) B(0) \rangle=\langle C(t) \rangle +  i \langle D(t) \rangle$. According to the quantum mechanical postulates, there exist two measurement apparatus associated with observables $C(t)$ and $D(t)$.  In this way, we may formally compute $\langle A(t) B(0) \rangle$ from the measured $\langle C(t) \rangle$ and $\langle D(t) \rangle$. However, the determination of $\langle C(t) \rangle$ and $\langle D(t) \rangle$ depends nontrivially on the correlation times and on the complexity of the specific time-evolution operator $U(t)$. Furthermore, we point out that the computation of $n$-time correlations, as $\langle \Psi | \Psi' \rangle = \langle \Psi | U^{\dagger}(t) A U(t) B | \Psi \rangle$, is not a trivial task even if one has access to full state tomography, due to the ambiguity of the global phase of state $| \Psi' \rangle = U^{\dagger}(t) A U(t) B | \Psi \rangle$. Therefore, we are confronted with a cumbersome problem: the design of measurement apparatus depending on the system evolution for determining $n$-time correlations of a system whose evolution may not be accessible. To our knowledge, a general formalism to attack this problem is still missing, while alternative algorithmic strategies~\cite{Mosca} may be considered.

In this Letter, we propose an efficient quantum algorithm for computing general \(n\)-time correlation functions of an arbitrary quantum system, requiring only an initially added probe and control qubit.  Moreover, our method is applicable to a general class of interacting spinorial, bosonic, and fermionic systems. Finally, we provide examples of our protocol in the frame of the linear response theory, where \(n\)-time correlation functions are needed. 

\begin{figure}[t]
\begin{center}
\vspace{0.5cm}
\hspace{-0.3cm}
\includegraphics [width= 1.02 \columnwidth]{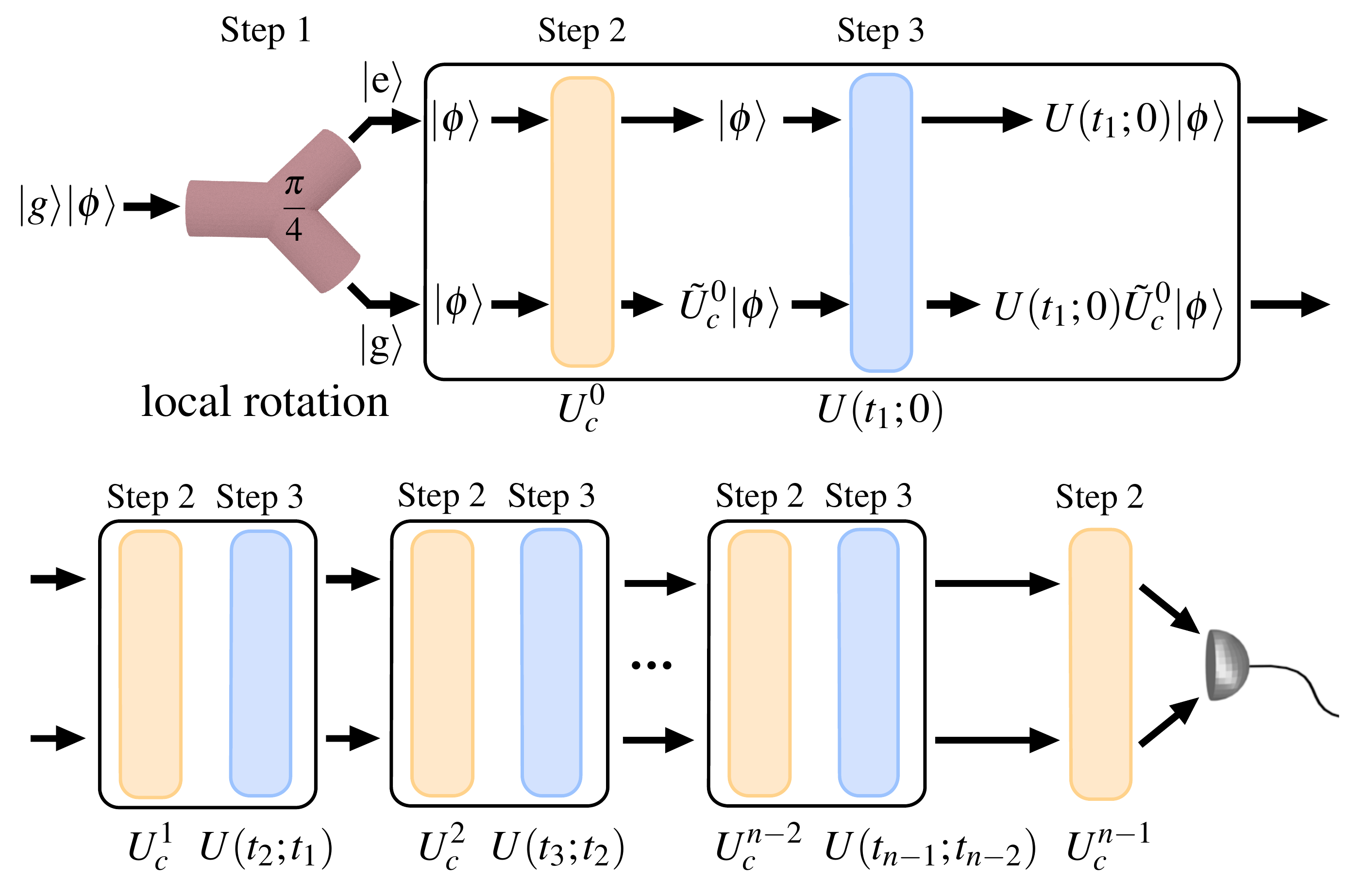}
\end{center}
\caption{(color online) Quantum algorithm for computing $n$-time correlation functions. The ancilla state $\frac{1}{\sqrt 2}(|{ e}\rangle + |{ g }\rangle)$ generates the $|{ e}\rangle$ and $|{  g }\rangle$ paths, step $1$, for the ancilla-system coupling. After that, controlled gates $U_c^m$ and unitary evolutions $U(t_m; t_{m-1})$ applied to our system, steps $2$ and $3$,  produce the final state $\Phi$. Finally, the measurement of the ancillary spin operators $\sigma_x$ and $ \sigma_y$ leads us to \(n\)-time correlation functions.}\label{figscheme}
\end{figure}

The protocol works under the following two assumptions.  First, we are provided with a controllable quantum system undergoing a given quantum evolution described by the Schr\"odinger equation
\begin{equation}\label{scho}
i\hbar\partial_t |\phi\rangle = H |\phi\rangle.
\end{equation}
Second, we require the ability to perform entangling operations, for example M\o lmer-S\o rensen~\cite{Sorensen99}  or equivalent controlled gates~\cite{Nielsen},  between  some part of the system and the ancillary qubit. More specifically, and as it is discussed in the Supplemental Material~\cite{Suplemental}, we require a number of entangling gates that grows linearly  with the order $n$ of the $n$-time correlation function and that remains fixed with increasing system size. This protocol  will provide us with  the efficient measurement of  generalized \(n\)-time correlation functions  of the form ${\langle\phi| O_{n-1}(t_{n-1}) O_{n-2}(t_{n-2}),...,O_1(t_1) O_0(t_0)|\phi\rangle}$, where $O_{n-1}(t_{n-1}),...,O_0(t_0)$ are certain operators evaluated at different times, e.g., $O_k(t_k) = U^{\dag}(t_k; t_0) O_k \ U(t_k; t_0)$, $U(t_k; t_0)$ being the unitary operator evolving the system from $t_0$ to $t_k$. For the case of dynamics governed by time-independent Hamiltonians, $U(t_k; t_0)=U(t_k - t_0) = e^{-(i/\hbar)H(t_k - t_0)}$. However, our method applies also to the case where $H=H(t)$, and can be sketched as follows. First, the ancillary qubit  is prepared in state ${\frac{1}{\sqrt 2}(|{ e} \rangle + |{ g } \rangle)}$ with ${|{  g }\rangle}$ its ground state, as in step $1$ of Fig.~\ref{figscheme}, so that the whole ancilla-system quantum state is ${\frac{1}{\sqrt 2}(|{ e} \rangle + |{ g }\rangle) \otimes |\phi\rangle}$, where $|\phi\rangle$ is the  state of the system. Second, we apply  the controlled quantum gate ${U^0_c = \exp{[ - (i / \hbar) |{ \rm g }\rangle\langle { \rm g }| \otimes  H_0 \tau_0 ] }}$, where, as we will see below, $H_0$ is a Hamiltonian related to the operator $O_0$, and $\tau_0$ is the gate time. As we point out in the Supplemental Material \cite{Suplemental}, this entangling gate can be implemented efficiently with M\o lmer-S\o rensen gates for operators $O_0$ that consist of a tensor product of Pauli matrices~\cite{Sorensen99}. This operation entangles the ancilla with the system generating the state $ \frac{1}{\sqrt 2} (|{ e}\rangle \otimes |\phi\rangle + |{ g }\rangle \otimes \tilde{U}_c^0|\phi\rangle)$, with $\tilde{U}_c^0= e^{- (i / \hbar) H_0 \tau_0}$,  step $2$ in Fig.~\ref{figscheme}. Next, we switch on the dynamics of the system governed by  Eq.~(\ref{scho}). For the sake of simplicity let us assume $t_0=0$. The effect on the ancilla-system wave function is to produce the state ${\frac{1}{\sqrt{2}}\big(|{ e} \rangle \otimes U(t_1; 0) |\phi \rangle + |{  g }\rangle \otimes U(t_1; 0) \tilde{U}_c^0 |\phi\rangle \big)}$, step $3$ in Fig.~\ref{figscheme}. Note that, remarkably, this last step does not require an interaction between the system and the ancillary-qubit degrees of freedom nor any knowledge of the Hamiltonian $H$. 
These techniques, as will be evident below, will find a natural playground in the context of quantum simulations, preserving its analogue or digital character. If we iterate $n$ times step~$2$ and step~$3$ with a suitable choice of  gates and evolution times, we obtain the state $\Phi = \frac{1}{\sqrt{2}}(|{ e}\rangle \otimes U(t_{n-1}; 0 )  |\phi\rangle + |{ g }\rangle   \otimes \tilde{U}_c^{n-1}   U(t_{n-1}; t_{n-2}),...,U(t_{2}; t_{1})  \tilde{U}_c^1 U(t_1; 0)   \tilde{U}_c^0 |\phi\rangle )$. Now, we target the quantity  ${\rm Tr}( |{ e}\rangle\langle { g } |  | \Phi\rangle\langle\Phi|)$ by measuring the $\langle \sigma_x \rangle$ and $\langle \sigma_y \rangle$ corresponding to the ancillary degrees of freedom. Simple algebra leads us to 
\begin{widetext}
\begin{equation}
\label{near}
{\rm Tr}( |{\rm e} \rangle\langle { \rm g } |  | \Phi\rangle\langle\Phi|) = \frac{1}{2}\left(\langle\Phi|\sigma_x| \Phi\rangle + i \langle\Phi|\sigma_y| \Phi\rangle\right)  
\frac{1}{2}\langle\phi| U^{\dag}(t_{n-1}; 0)  \tilde{U}_c^{n-1}   U(t_{n-1}; t_{n-2}),...,U(t_2; t_1)  \tilde{U}_c^1 U(t_1; 0)   \tilde{U}_c^0 |\phi\rangle.  
\end{equation}
\end{widetext}
It is easy to see that,  by using the composition property ${U(t_k; t_{k-1}) = U(t_k; 0) U^{\dag}(t_{k-1}; 0)}$, Eq.~(\ref{near}) corresponds to a general construction relating $n$-time correlations of system operators $\tilde{U}_c^{k}$ with two one-time ancilla measurements. In order to explore its depth, we shall examine several classes of systems and suggest concrete realizations of the proposed algorithm. The crucial point is  establishing a connection that associates the $\tilde{U}_c^k$ unitaries with $O_k$ operators. 

Starting with  the discrete variable case, e.g., spin systems, and profiting from the fact that Pauli matrices are both Hermitian and unitary, it follows that
\begin{equation}
\label{linear}
\tilde{U}_c^m\big|_{\Omega \tau_m =\pi/2 }= \exp{[ - (i / \hbar) H_m \tau_m]}\big|_{\Omega\tau_m =\pi/2 } = -i O_m,
\end{equation}
where $H_m = \hbar \Omega O_m$, $\Omega$ is a coupling constant, and  $O_m$ is a tensor product of Pauli matrices of the form ${O_m = \sigma_{i_m} \otimes \sigma_{j_m},...,\sigma_{k_m}}$ with $i_m, j_m, ..., k_m \in 0, x, y, z$, and $\sigma_0=\mathbb{I}$. 
In consequence, the controlled quantum gates in step 2 correspond to $U_c^m\big|_{\Omega \tau_m =\pi/2 }= \exp{(-i |{ g }\rangle\langle { g } | \otimes \Omega O_m \tau_m)}$, which can be implemented efficiently, up to local rotations, with four M\o lmer-S\o rensen gates~\cite{Sorensen99, Mueller11, Lanyon11, Casanova12}. In this way, we can write the second line of Eq.~(\ref{near}) as 
\begin{eqnarray}\label{ntimespin}
(-i)^{n}\langle\phi| O_{n-1}(t_{n-1}) O_{n-2}(t_{n-2}) ... O_0(0)|\phi\rangle,
\end{eqnarray}
which amounts to the measured \( n \)-time correlation function of Hermitian and unitary operators $O_k$. We can also apply these ideas to the case of non-Hermitian operators, independent of their unitary character, by considering  linear superpositions of the Hermitian objects appearing in Eq.~(\ref{ntimespin}).

We show now how to apply this result to the case of fermionic systems. In principle, the previous proposed steps would apply straightforwardly if we had access to the corresponding fermionic operations. In the case of quantum simulations, a similar result is obtained via the Jordan-Wigner mapping of fermionic operators to tensorial products of Pauli matrices, $b^\dag_p\rightarrow\Pi_{r=1}^{p-1}\sigma_{+}^p\sigma_z^r$~\cite{ JordanWigner}. Here, $b^{\dag}_p$ and $b_q$ are creation and annihilation fermionic operators obeying anticommutation relations, $\{b^{\dag}_p, b_q \} = \delta_{p, q}$. For trapped ions, a quantum algorithm for the efficient implementation of fermionic models has recently been proposed~\cite{Casanova12, Lamata13, Yung13}. 
Then, we code
$\langle b_p^{\dag}(t) b_q(0)\rangle = \langle\Phi|(\sigma^p_+\otimes\sigma^{p-1}_z,...,\sigma_z^1)_{t} \  \sigma^q_-\otimes\sigma^{q-1}_z,...,\sigma_z^1|\Phi\rangle$, where 
$(\sigma^p_+\otimes\sigma^{p-1}_z,...,\sigma_z^1)_{t} = e^{\frac{i}{\hbar}H t} \sigma^p_+\otimes\sigma^{p-1}_z,...,\sigma_z^1 e^{- (i / \hbar) H t}$. Now, taking into account that $\sigma_{\pm} = \frac{1}{2}(\sigma_x \pm i \sigma_y)$, the fermionic correlator $\langle b_p^{\dag}(t) b_q(0)\rangle$ can be written as the sum of four terms of the kind appearing in Eq.~(\ref{ntimespin}). This result extends naturally to multitime correlations of fermionic systems.

The case of bosonic $n$-time correlators requires a variant in the proposed method, due to the nonunitary character of the associated bosonic operators. In this sense, to reproduce a linearization similar to that of Eq.~(\ref{linear}), we can write
\begin{equation}
\partial_{\Omega \tau_m}\tilde{U}_c^m\big|_{\Omega \tau_m =0} =\partial_{\Omega \tau_m}\exp{[- (i / \hbar) H_m \tau_m]}\big|_{\Omega\tau_m =0 } =-i O_m,
\end{equation}
with $H_m = \hbar \Omega O_m$. Consequently, it follows that
\begin{eqnarray}
\label{nolabel}
&&\partial_{\Omega \tau_j},...,\partial_{\Omega\tau_k} {\rm Tr}( |{\rm e}\rangle\langle { g } |  | \Phi\rangle\langle\Phi|)\big|_{\Omega(\tau_\alpha ,..., \tau_\beta) = \pi /2, \ \Omega(\tau_j ,..., \tau_k) = 0} =\nonumber\\
&&\ \ \ \ \ (-i)^{n}\langle\phi| O_{n-1}(t_{n-1}) O_{n-2}(t_{n-2}) ,..., O_0(0)|\phi\rangle \, ,
\end{eqnarray}
where the label $(\alpha, ..., \beta)$ corresponds to spin operators and $(j,...,k)$ to spin-boson operators. The right-hand side is a correlation of Hermitian operators, thus substantially extending our previous results. For example, $O_m$ would include spin-boson couplings as $O_m = \sigma_{i_m} \otimes \sigma_{j_m},...,\sigma_{k_m} (a + a^{\dag})$. The way of generating the associated evolution operator ${\tilde{U}_c^m = \exp(-i \Omega O_m \tau_m)}$ has been shown in~\cite{Casanova12, Lamata13, Mezzacapo12}, see also the Supplemental Material~\cite{Suplemental}. Note that, in general, dealing with discrete derivatives of experimental data is an involved task~\cite{Lougovsky06, CasanovaEPJD}. However, recent experiments in trapped ions~\cite{Gerritsma1, Gerritsma2, Zahringer} have already succeeded in the extraction of precise information from data associated to first- and second-order derivatives. 

The method presented here works as well when the system is prepared in a mixed-state $\rho_0$, e.g. a state in thermal equilibrium~\cite{Kubo57, Zwanzig65}.  Accordingly, for the case of spin correlations, we have 
\begin{eqnarray}
\nonumber
{\rm Tr}(|{e}\rangle\langle { g } | \tilde{\rho}) =(-i)^{n}{\rm Tr}( O_{n-1}(t_{n-1})O_{n-2}(t_{n-2}) ,..., O_0(0) \rho_0) , \\
\end{eqnarray}
with
\begin{eqnarray}
\tilde{\rho} =  \big[... U(t_2; t_1)U_c^1U(t_1; 0)U_c^0 \big]  \tilde{\rho}_0  \big[ U_c^{0 \dag} U(t_1; 0)^{\dag} U_c^{1\dag}U(t_2; t_1)^{\dag} ... \big] \nonumber \\
\end{eqnarray}
and $\tilde{\rho}_0 = \frac{1}{2}(|{ e}\rangle+|{ g } \rangle)(\langle {e}| + \langle {g } |)\otimes \rho_0$.
If bosonic variables are involved, the analogue to Eq.~({\ref{nolabel}) reads
\begin{eqnarray}
&&\partial_{\Omega \tau_j} ,..., \partial_{\Omega\tau_k}{\rm Tr}(|{e}\rangle\langle {g } | \tilde{\rho})\big|_{\Omega(\tau_\alpha ,..., \tau_\beta) = \pi / 2, \ \Omega(\tau_j ,..., \tau_k) = 0} =\nonumber\\&&\ \ \ \ \ \ \ (-i)^n{\rm Tr}( O_{n-1}(t_{n-1})O_{n-2}(t_{n-2}) ,..., O_0(0) \rho_0) .
\end{eqnarray}

We will exemplify the introduced formalism with the case of quantum computing of spin-spin correlations of the form 
\begin{equation}\label{ss}
\langle\sigma_i^k(t) \sigma_{j}^{l}(0)\rangle,
\end{equation}
where $k, l = x, y, z$, and $i,j = 1,... ,N$, $N$ being the number of spin particles involved. In the context of spin lattices, where several quantum models can be simulated in different quantum platforms as trapped ions~\cite{Britton12, Porras04, Friedenauer08, Kim09, Kim10, Richerme13}, optical lattices~\cite{Ripoll04, Simon11, Greiner02}, and circuit QED~\cite{Ripoll08, Tian10, Zang13, Viehmann13},  correlations like~(\ref{ss}) are a crucial element in the computation of, for example, the magnetic susceptibility~\cite{Kubo57, Zwanzig65, Forster}. In particular, with our protocol, we have access to  the frequency-dependent susceptibility  $\chi_{\sigma, \sigma}^\omega$
that quantifies the linear response of a spin system when it is driven by a monochromatic field. This situation is described by  the Schr\"odinger equation  $i\hbar \partial_t | \psi \rangle = (H + f_{\omega} \sigma_j^l e^{i\omega t}) | \psi \rangle$, where, for simplicity, we assume $H\neq H(t)$. With a perturbative approach, and following the Kubo relations~\cite{Kubo57, Zwanzig65}, one can calculate the first-order effect of a magnetic  perturbation acting on the  \( j \)th spin in the polarization of the \( i \)th spin as
\begin{equation}\label{lr}
\langle\sigma_i^{k} (t) \rangle = \langle\sigma_i^{k} (t) \rangle_0+\chi_{\sigma, \sigma}^\omega  f_\omega e^{i\omega t}.
\end{equation}
Here, $\langle\sigma_i^{k} (t) \rangle_0$ corresponds to the value of  the observable $\sigma_i^k$ in the absence of perturbation, and the frequency-dependent susceptibility $\chi_{\sigma, \sigma}^\omega$ is
\begin{equation}
\chi_{\sigma, \sigma}^\omega=\int_0^t ds \ \phi_{\sigma, \sigma}(t-s) e^{i\omega (s-t)}
\end{equation}
where $ \phi_{\sigma, \sigma}(t-s)$ is called the response function, which can be written in terms of two-time correlation functions
\begin{eqnarray}\label{aef}
\phi_{\sigma, \sigma}(t-s) &=& (i / \hbar) \langle [\sigma_i^k(t-s),\sigma_j^l(0)]\rangle \nonumber \\
&=&  (i / \hbar){\rm Tr}\big( [\sigma_i^k(t-s),\sigma_j^l(0)] \rho\big), 
\end{eqnarray}
with $\rho = U(t)\rho_0 U^{\dag}(t)$, $\rho_0$ being the initial state of the system and $U(t)$ the perturbation-free time-evolution operator ~\cite{Kubo57}. Note that for thermal states or energy eigenstates, we have  $\rho = \rho_0$. According to our proposed method, and assuming for the sake of simplicity $\rho=|\Phi\rangle\langle\Phi|$, the measurement of the commutator in Eq.~(\ref{aef}), corresponding to the imaginary part of ${\langle \sigma_i^k(t-s) \sigma_j^l(0) \rangle }$,  would require the following  sequence of interactions: ${| \Phi \rangle \rightarrow U_c^1 U(t-s)U_c^0 | \Phi \rangle}$, where ${U_c^0=e^{-i | {  g } \rangle \langle { g } | \otimes \sigma_j^l \Omega \tau}}$, ${U(t-s)=e^{ -(i / \hbar) H (t-s)}}$, and ${U_c^1=e^{-i | { g } \rangle \langle { g } | \otimes \sigma_i^k \Omega \tau}}$, for $\Omega \tau = \pi /2$. After such a gate sequence, the expected value in Eq. (\ref{aef}) corresponds to $-1/2  \langle \Phi | \sigma_y | \Phi \rangle $. In the same way, Kubo relations allow the computation of higher-order corrections of the perturbed dynamics in terms of higher-order  time correlation functions. In particular, second-order corrections to the linear response of Eq.~(\ref{lr}) can be calculated through the computation of three-time correlation functions of the form $\langle \sigma_i^k(t_2) \sigma_j^l(t_1) \sigma_j^l(0)\rangle$. Using the method introduced in this paper, to measure such a three-time correlation function one should perform the evolution $| \Phi \rangle \rightarrow U_c^1U(t_2 - t_1)U_c^0 U(t_1)U_c^0 | \Phi \rangle$, where $U_c^0=e^{-i | { g } \rangle \langle { g } | \otimes \sigma_j^l \Omega \tau}$, $U(t)=e^{- (i / \hbar) H t}$ and $U_c^1=e^{-i | { g } \rangle \langle { g } | \otimes \sigma_i^k \Omega \tau}$ for $\Omega \tau = \pi /2$. The searched time correlation then corresponds to the quantity $1/2(i \langle \Phi | \sigma_x | \Phi \rangle -  \langle \Phi | \sigma_y | \Phi \rangle  )$.

Our method is not restricted to corrections of observables that involve the spinorial degree of freedom. Indeed, we can show how the method applies when one is interested in the effect of the perturbation onto the motional degrees of freedom of the involved particles. According to the linear response theory, corrections to observables involving the motional degree of freedom enter in the response function, ${\phi_{a+a^{\dag}, \sigma}(t-s)}$, as time correlations of the type $\langle  (a_i + a^{\dag}_i)_{(t-s)}  \sigma_j^l\rangle$, where ${(a_i + a^{\dag}_i)_{(t-s)}=e^{(i / \hbar)H(t-s)}(a_i + a^{\dag}_i) e^{-(i / \hbar) H(t-s)}}$. The response function can be written as in Eq.~(\ref{aef}) but replacing the operator $\sigma_i^k(t-s)$ with $(a_i + a^{\dag}_i)_{(t-s)}$.
The corrected expectation value is now
\begin{equation}\label{lrb}
\langle(a_i + a_i^{\dag})_{t} \rangle = \langle (a_i + a_i^{\dag})_{t} \rangle_0+\chi_{a+a^{\dag}, \sigma}^\omega  f_\omega e^{i\omega t}.
\end{equation}
In this case, the gate sequence for the measurement of the associated correlation function $\langle  (a_i + a^{\dag}_i)_{(t-s)}  \sigma_j^l\rangle$ reads $| \Phi \rangle \rightarrow U_c^1 U(t-s)U_c^0 | \Phi \rangle$, where ${U_c^0=e^{-i | {g } \rangle \langle {g } | \otimes \sigma_j^l \Omega_0 \tau_0}}$, ${U(t-s)=e^{- (i / \hbar) H (t-s)}}$, and ${U_c^1=e^{-i | { g } \rangle \langle { g } | \otimes (a_i + a^{\dag}_i) \Omega_1 \tau_1}}$, for ${\Omega_0 \tau_0 = \pi /2}$. The time correlation is now obtained through the first derivative $ -1/2 \partial_{\Omega_1 \tau_1} (\langle \Phi | \sigma_x | \Phi \rangle + i  \langle \Phi | \sigma_y | \Phi \rangle) |_{\Omega_1 \tau_1 = 0}$.

Equations~(\ref{lr}) and (\ref{lrb}) can be extended to describe the effect on the system of light pulses containing frequencies in a certain interval $(\omega_0, \omega_0+\delta)$. In this case, Eqs.~(\ref{lr}) and (\ref{lrb}) read}
\begin{equation}\label{lrmf}
\langle\sigma_i^{k} (t) \rangle = \langle\sigma_i^{k} (t) \rangle_0+\int_{\omega_0}^{\omega_0 + \delta}\chi_{\sigma, \sigma}^\omega  f_\omega e^{i\omega t} d\omega,
\end{equation}
and  
\begin{equation}\label{lrbmf}
\langle(a_i + a_i^{\dag})_{t} \rangle = \langle (a_i + a_i^{\dag})_{t} \rangle_0 + \int_{\omega_0}^{\omega_0+\delta}\chi_{a+a^{\dag}, \sigma}^\omega  f_\omega e^{i\omega t} d\omega.
\end{equation}
Note that despite the presence of many frequency components of the light field in the integrals of Eqs.~(\ref{lrmf}) and  (\ref{lrbmf}), the computation of the susceptibilities, $\chi_{\sigma, \sigma}^\omega $ and $\chi_{a+a^{\dag}, \sigma}^\omega$, only requires the knowledge of the time correlation functions $\langle [\sigma_i^k(t-s),\sigma_j^l(0)]\rangle$ and $\langle [(a+a^{\dag})_{(t-s)},\sigma_j^l(0)]\rangle$, which can be efficiently calculated with the protocol described in  Fig~\ref{figscheme}. In this manner, we provide an efficient quantum algorithm to characterize the response of different quantum systems to external perturbations. Our method may be related to the quantum computation of transition probabilities ${|\alpha_{{ f},{ i}}(t)|^2=|\langle {  f }| U(t) | { i } \rangle|^2=\langle { i} | P_{ f} (t) | { i} \rangle }$, between initial and final states,  $ | {i} \rangle$ and $| {f} \rangle$, with $P_{ f} (t)=U(t)^{\dag}| {f}\rangle\langle { f} | U(t)$, and to transition or decay rates $\partial_t |\alpha_{{ f},{ i }}(t)|^2$ in atomic ensembles. These questions are of general interest for evolutions perturbed by external driving fields or by interactions with other quantum particles. 

In conclusion, we have presented a quantum algorithm to efficiently compute  arbitrary $n$-time correlation functions. The protocol requires the initial addition of a single probe and control qubit and is valid for arbitrary unitary evolutions. Furthermore, we have applied this method to interacting fermionic, spinorial, and bosonic systems,  showing how to compute second-order effects beyond the linear response theory. Moreover, if used in a quantum simulation, the  protocol preserves the analogue or digital character of the associated dynamics. We believe that the proposed concepts pave the way for making accessible a wide class of $n$-time correlators in a wide variety of physical systems. 

The authors acknowledge support from Spanish MINECO FIS2012-36673-C03-02, UPV/EHU UFI 11/55, UPV/EHU PhD grant, Basque Government IT559-10 and IT472-10, and CCQED, PROMISCE, and SCALEQIT European projects.

\begin{widetext}

\section{Supplemental Material for \\ ``Efficient Quantum Algorithm for Computing n-time Correlation Functions''}

In this suplemental material we provide additional discussions about the efficiency of our method together with comparisons of our method to existing ones and specific calculations concerning the implementation of gates. 

\section{I. Efficiency of the method}

Our algorithm is conceived to be run in a setup composed of a system undergoing the evolution of interest and an ancillary qubit. Thus, the size of the setup where the algorithm is to be run is always that of the system plus one qubit, regardless of the order of the time correlation. For instance, if we are considering an $N$-qubit system then our method is performed in a setup composed of $N+1$ qubits. 

With respect to time-efficiency,  our algorithm requires the performance of $n$ controlled gates $U_c^i$ and $n-1$  time-evolution operators $U(t_{j+1}; t_{j})$, $n$ being  the order of the time-correlation function.  

If we assume that $q$ gates are needed for the implementation of the system evolution and $m$ gates are required per control operation, our algorithm employs $(m+q)*n - q$ gates. As $m$ and $q$ do not depend on the order of the time correlation we can state that our algorithm needs a number of gates that scales  as a first-order polynomial with respect to the order $n$ of the time-correlation function. 

The scaling of $q$ with respect to the  system size depends on the specific simulation  under study. However, for most relevant cases it can be shown that this scaling is polynomial. For instance, in the case of an analogue quantum simulation of unitary dynamics, what it is usually called an always-on simulation, we have $q=1$. For a model requiring digital techniques $q$ will scale polinomially if the number of terms in the Hamiltonian grows polinomially  with the number of constituents, which is a physically-reasonable assumption~\cite{Lloyd96, Nielsen, Casanova12}. In any case, we want to point out that the way in which $q$ scales is a condition inherent to any quantum simulation process, and, hence, it is not an additional overhead introduced by our proposal.

With respect to the number $m$, and as  explained in the next section, this number does not depend on the system size, thus, from the point of view of efficiency  it amounts to  a constant factor. 

In order to provide a complete runtime analysis of our protocol we study now the number of iterations needed to achieve a certain precision $\delta$ in the measurement of the time correlations. According to Berntein's inequality~\cite{bernstein} we have that 
\begin{equation}
\Pr\left[\left|\frac{1}{L}\sum_{i=1}^LX_i-\langle X\rangle\right|>\delta\right]\leq2 \exp \left( \frac{-L \delta^2}{4 \sigma_0^2} \right),
\end{equation}
where $X_i$ are independent random variables, and $\sigma_0^2$ is a bound on their variance. Interpreting $X_i$ as a single observation of the real or imaginary part of the time-correlation function, we find the number of measurements needed to have a precision $\delta$. Indeed, we have that $\left|\frac{1}{L}\sum_{i=1}^LX_i-\langle X\rangle\right|\leq\delta$ with probability $P\geq 1-e^{-c}$, provided that $L\geq \frac{4(1+c)}{\delta^2}$, where we have set $\sigma_0^2 \le 1$, as we always measure Pauli observables. This implies that the number of gates that we need to implement to achieve a precision $\delta$ for the real or the imaginary part of the time-correlation function is $\frac{4(1+c)}{\delta^2} [(m+q)n - q]$. Again, $c$ is a constant factor which does not depend nor on the order of the time correlation neither on the size of the system.

\section{II. N-body interactions  with M\o lmer-S\o rensen gates}\label{SecMolmer}

Exponentials of tensor products of Pauli operators, $\exp[i\phi\sigma_1 \otimes \sigma_2 \otimes ... \otimes \sigma_k]$, can be systematically constructed, up to local rotations, with a M\o lmer-S\o rensen gate applied over the $k$ qubits, one local gate on one of the qubits, and the inverse M\o lmer-S\o rensen gate on the whole register. This can be schematized as follows,
\begin{equation}
U = U_{MS}(-\pi/2, 0) U_{\sigma_z} (\phi) U_{MS}(\pi/2, 0) = \exp [ i\phi \sigma_1^z \otimes \sigma_2^x \otimes ...\otimes \sigma_k^z],
\end{equation}
where $U_{MS}(\theta, \phi ) = \exp[-i\theta ( \cos \phi S_x + \sin \phi S_y)^2/4]$, $S_{x,y} = \sum_{i=1}^k \sigma_i^{x,y}$ and $U_{\sigma_z}(\phi) = \exp(i \phi' \sigma_1^z)$ for odd $k$, where $\phi ' = \phi $ for $k=4n+1$, and $\phi '= - \phi $ for $k=4n-1$, with positive integer n. For even $k$, $U_{\sigma_z}(\phi)$ is replaced by $U_{\sigma_y}(\phi) = \exp(i\phi' \sigma_1^y)$, where $\phi '=\phi$ for $k=4n$, and $ \phi ' = \phi$ for $k=4n-2$, with positive integer $n$. Subsequent local rotations will generate any combination of Pauli matrices in the exponential. 

The replacement in the previous scheme of the central gate  $U_{\sigma_z}(\phi)$ by an interaction containing a coupling with bosonic degrees of freedom, for example  $U_{\sigma_z , (a+a^{\dag})}(\phi) = \exp[i \phi' \sigma_1^z (a +a^{\dag}) ] $, will directly provide us with 
\begin{equation}\label{interaction}
U = U_{MS}(-\pi/2, 0) U_{\sigma_z, (a+a^\dag)} (\phi) U_{MS}(\pi/2, 0) = \exp [ i\phi \sigma_1^z \otimes \sigma_2^x \otimes ...\otimes \sigma_k^z (a+a^{\dag})].
\end{equation}

In order to provide a complete recipe for systems where M\o lmer-S\o rensen interactions are not directly available, we want to comment that the kind of entangling quantum gates required by our algorithm, see the right hand side of Eq.~(\ref{interaction}) above, are always decomposable in a polynomial sequence of controlled-Z  gates~\cite{Nielsen}. For example, in the case of a three-qubit system we have  
\begin{equation}
CZ_{1,3} CZ_{1,2} e^{-i\phi\sigma^y_1}CZ_{1,2} CZ_{1,3} = \exp{(-i\phi \sigma^y_1\otimes \sigma^z_2 \otimes \sigma^z_3)}
\end{equation}
Here, $CZ_{i,j}$ is a controlled-Z gate between the $i, j$ qubits and $e^{-i\phi\sigma^y_1}$ is local rotation applied on the first qubit. This result can be easily extended to $n$-qubit systems with the application of $2(n-1)$ controlled operations~\cite{Nielsen}.

Therefore, it is demonstrated the polynomial character of our algorithm, and hence its efficiency, even if M\o lmer-S\o rensen gates are not available in our setup.

\section{III. Our protocol VS Hadamard and SWAP tests}

Two typical approaches for the measurement of correlations in the context of quantum computing are the Hadamard and the SWAP tests. The Hadamard test is performed in a setup consisting of the system of interest and a qubit, and thus in terms of space is as efficient as our algorithm. The performance of a Hadamard gate followed by a controlled unitary evolution, another Hadamard gate and the measurement of two ancilla operators will lead to the real and imaginary parts of a correlation of the type $\langle U \rangle$, where $U$ corresponds to the controlled unitary. While the evolutions of the system of interest are not controlled in our protocol, the Hadamard test needs to perform control unitary operations which may not be trivial for many body Hamiltonians or Hamiltonians depending in time. In this sense our method supposes a significant step forward in simplicity and a notable reduction in the requirements of our setup. It is noteworthy to mention that our algorithm could access time correlations of systems that undergo non-unitary dynamics.

In the case of the SWAP test, correlations between two states are measured following a similar scheme, in this case a Hadamard gate is performed on the ancilla qubit, after that a control SWAP gate is implemented between the two states of interest and finally a second Hadamard gate is performed on the ancillary qubit. Again local measurements on the ancillary qubit will provide real and imaginary parts of the correlation between the two states. While our protocol involves only one ancillary qubit, $N+1$, the SWAP test needs two copies of the system and the ancillary qubit which makes a total of $2N+1$ qubits, this makes our protocol significantly more space saving.

\end{widetext}

\end{document}